# The Definition of Quasicrystals

RON LIFSHITZ

School of Physics & Astronomy, Raymond and Beverly Sackler Faculty of Exact Sciences, Tel Aviv University, Tel Aviv 69978, Israel.

**Abstract**

It is argued that the definition of quasicrystals should not include the requirement that they possess an axis of symmetry that is forbidden in periodic crystals. The term "quasicrystal" should simply be regarded as an abbreviation for "quasiperiodic crystal," possibly with two provisos, as discussed below. The argument is supported by theoretical as well as experimental examples of quasicrystals without any forbidden symmetry.

## 1. Introduction

The aim of this paper is to argue against the common practice[1-3] to restrict the definition of quasicrystals by requiring that they possess an axis of symmetry that is forbidden in periodic crystals. According to this restriction there are no quasicrystals in 1-dimension, and a quasicrystal in 2- or 3-dimensions must have an axis of $N$-fold symmetry, with $N=5$, or $N>6$. In this paper I urge the scientific community to accept the original definition of Levine and Steinhardt[4] whereby the term "quasicrystal" is simply an abbreviation for "quasiperiodic crystal," possibly with two provisos: (a) that the quasicrystal is strictly aperiodic (because the mathematical definition of quasiperiodicity includes periodicity as a special case); and (b) that its diffraction diagram contains no clear subset of strong Bragg peaks—"main reflections"—accompanied by weak peaks—"satellites"—to distinguish it from an incommensurately modulated crystal and an incommensurate composite crystal (a point of view, promoted, for example, by Janssen,[5] and in somewhat more technical terms by Katz and Gratias[6]).

I shall first review the definitions of the terms mentioned in the previous paragraph. I will then refer to experimental observations as well as tiling models of structures which should be considered as quasicrystals even though they possess no forbidden symmetries. Examples of square quasicrystals are shown in Figure 1.

## 2. What is a crystal?

Before Shechtman's 1982 discovery of the first icosahedral quasicrystal[7] long-range order was thought to be synonymous with periodicity. Crystals, or ordered solids, were in fact defined as being periodic arrangements of atoms, with the allowance of certain modifications to the underlying periodicity as in the cases of incommensurately modulated crystals and incommensurate composite crystals. Nearly two decades later, it is now well established that periodicity is not necessary for producing long-range order, but it is still not quite clear how exactly one should characterize the existence of order.

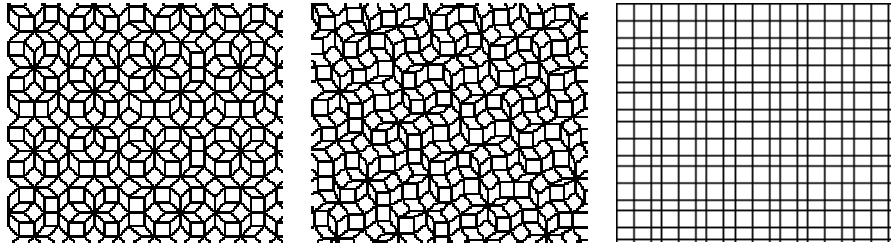

**Figure 1**. Three different quasicrystals: (Left) A quasiperiodic tiling with 8-fold symmetry. (Center) A quasiperiodic tiling with 4-fold symmetry, created by distorting the 8-fold tiling. (Right) A quasiperiodic 4-fold tiling, created by a 90 degree superposition of two Fibonacci grids.

Noting that the diffraction spectra of all experimentally observed crystals contain Bragg peaks, the International Union of Crystallography[8] has redefined a crystal to be "…*any solid having an essentially discrete diffraction diagram.*" Crystals that are periodic on the atomic scale are explicitly called *periodic crystals*, all others are called *aperiodic crystals*.

This definition is not merely an empirical one but is also motivated by the common practice of taking the set of density Fourier coefficients $\rho(\mathbf{k})$ at non-zero wave vectors $\mathbf{k}$ as the Landau order parameter, signaling a phase transition from a liquid state to an ordered solid state. The new definition was intentionally made vague by the inclusion of the word "essentially." It is meant only as a temporary working-definition until a better understanding of crystallinity emerges. We need not worry about this vagueness because we shall only be concerned with quasiperiodic crystals, which are a well defined subcategory of structures, satisfying the new definition.

### 3. What is a quasiperiodic crystal?

Crystals whose density functions may be expanded as a superposition of a countable number of plane waves are called *almost periodic crystals*. In particular, if taking integral linear combinations of a finite number $D$ of wave vectors in this expansion can span all the rest, then the crystal is *quasiperiodic*. Each diffraction peak is then indexed by $D$ integers. Periodic crystals are the special case where the indexing dimension $D$ is equal to the actual physical dimension of the crystal. All experimentally observed crystals to date are quasiperiodic.

Although an official nomenclature has not yet been agreed upon, one clearly identifies (at least) two special categories among the family of quasiperiodic crystals: incommensurately modulated crystals and incommensurate composite crystals. (For details see van Smaalen[2] and Lifshitz.[9])

*Incommensurately modulated crystals* consist of a basic (or average) ordered structure that is perturbed periodically—"modulated"—where the period of the modulation is incommensurate with the underlying periodicities of the basic structure. The diffraction diagrams of incommensurately modulated crystals are characterized by having a subset of "main reflections"—Bragg peaks which are significantly brighter than the others—describing the basic structure, and a set of weaker peaks, called "satellites," arising from the modulation. The basic structure itself can be either periodic or not.

*Incommensurate composite crystals*, also called *intergrowth compounds*, are composed of two or more interpenetrating subsystems with mutually incommensurate periodicities. Each subsystem when viewed independently is itself a crystal—in all experimentally observed examples a periodic one, but in theory not necessarily so—which is incommensurately modulated due to its interaction with the other subsystems. The diffraction diagrams of composite crystals are characterized by the existence of two or more subsets of main reflections, caused by the average structures of the different subsystems, and a set of weak reflections caused by the modulations.

## 4. What is a quasicrystal?

There are also (strictly aperiodic) quasiperiodic crystals for which a description in terms of a modulation of a basic structure or a composition of two or more substructures is either inappropriate or impossible. I argue that one should refer to all such crystals as *quasicrystals*, regardless of their point-group symmetry. The most common model for such crystals is a quasiperiodic tiling such as the famous Penrose tiling. One fills space with "unit cells" or "tiles" in a way that maintains long-range order without periodicity, and produces an essentially discrete diffraction diagram.

Clearly, quasiperiodic crystals possessing symmetries that are forbidden for periodic crystals—such as the observed icosahedral, octagonal, decagonal, and dodecagonal crystals—cannot be formed by modifying an underlying periodic structure with the same symmetry, and are therefore all *quasicrystals*.[*] Quasiperiodic crystals with no forbidden symmetries can be formed as a modification of a periodic structure, but that need not be the case.

It turns out that there are experimentally observed quasiperiodic crystals with cubic symmetry,[11] as well as tetrahedral[12,13] tetragonal[14] and possibly also ones with with hexagonal[15] symmetry, that are neither modulated crystals nor composite crystals. Their diffraction diagrams show no clear subset(s) of main reflections, yet they do not possess any forbidden symmetry. These crystals are not formed by modifying an underlying periodic structure. They are as intrinsically quasiperiodic as the quasicrystals that have forbidden symmetries, and should therefore be considered quasicrystals as well.

From a theoretical standpoint it should be obvious that there is nothing special about point groups that are forbidden for periodic crystals. Any method that is used to generate a quasiperiodic tiling with, say, 10-fold symmetry can be used to generate quasiperiodic tilings with, say, 4-fold symmetry. Indeed, there are many examples in the literature of tiling models of quasicrystals, with 2-, 4-, and 6-fold symmetry, generated by all the standard methods: matching rules,[16] substitution rules,[16,17] the cut-and-project method[14,18] and by the dual grid method.[19] Figure 1 shows an 8-fold quasicrystal alongside two examples of 4-fold quasicrystals, generated using the dual grid method.

---

[*] There are also icosahedral quasicrystals that are incommensurately modulated, with diffraction peaks that are each indexed by 12 integers.[10]

In conclusion, quasicrystals in $d$-dimensional space can have any finite subgroup of $O(d)$ as their point group. It is only the introduction of $d$-dimensional periodicity that imposes restrictions on the allowed symmetry operations.